\documentclass[referee,sn-nature]{sn-jnl}

\pdfpagewidth=\paperwidth
\pdfpageheight=\paperheight

\usepackage{graphicx}%
\usepackage{amsmath,amssymb,amsfonts}%
\usepackage{bm}%
\usepackage{booktabs}%
\usepackage{xcolor}%
\usepackage{url}%
\usepackage{glossaries}%

\setacronymstyle{long-short}
\newacronym{loeo}{LOEO}{leave-one-element-out}
\newacronym{sinkhorn}{$S_{\varepsilon}$}{Sinkhorn divergence}
\newacronym{nsinkhorn}{$\widetilde{S}_{\varepsilon}$}{normalized Sinkhorn divergence}
\newacronym{umap}{UMAP}{Uniform Manifold Approximation and Projection}
\newglossaryentry{ef}{
    name={\ensuremath{f_X}},
    description={}
}
\newglossaryentry{efc}{
    name={\ensuremath{f_X^{\mathrm{c}}}},
    description={}
}
\newacronym{rf}{RF}{random forest}
\newacronym{xgb}{XGB}{extreme gradient boosting}
\newacronym{svr}{SVR}{radial-basis-function support-vector regression}
\newacronym{mae}{MAE}{mean absolute error}

\newacronym{engineered}{$\phi_{\mathrm{E}}$}{engineered descriptor}
\newacronym{pretrained}{$\phi_{\mathrm{PCG}}$}{pretrained crystal-graph encoder}

\begin{document}

\title{Materials representations organize transferable supervision for chemical extrapolation}

\author*[1,2]{\fnm{Enda} \sur{Xiao}}\email{Xiao.Enda@nims.go.jp}
\author*[1,2]{\fnm{Terumasa} \sur{Tadano}}\email{Tadano.Terumasa@nims.go.jp}

\affil[1]{\orgdiv{Research Center for Magnetic and Spintronic Materials},
\orgname{National Institute for Materials Science},
\orgaddress{\street{1-2-1 Sengen}, \city{Tsukuba}, \postcode{305-0047},
\state{Ibaraki}, \country{Japan}}}

\affil[2]{\orgdiv{Digital Transformation Initiative Center for Magnetic
Materials}, \orgname{National Institute for Materials Science},
\orgaddress{\street{1-2-1 Sengen}, \city{Tsukuba}, \postcode{305-0047},
\state{Ibaraki}, \country{Japan}}}

\abstract{Materials discovery often requires chemical extrapolation beyond
labeled data. When target compounds contain elements absent from training, which
compounds of other elements provide transferable supervision remains unclear.
Here we show that, without target-property labels, materials representations
rank other elements by the supervision their compounds provide. Engineered
descriptors rank them largely by elemental-property similarity, whereas a
pretrained crystal-graph encoder ranks them by the geometry of its learned
representation. At fixed training-set size, excluding the compounds of the
encoder's six top-ranked elements raises formation-energy and bandgap
errors by 100--110~meV atom$^{-1}$ and 42--85~meV, versus little change for six
lower-ranked elements. The rankings hold across properties, regressors and
datasets. The encoder's ranking outperforms elemental-property similarity and
periodic groups, suggesting that it captures how elements behave within
compounds. Materials representations thus organize transferable supervision,
which can be identified before model training, informing dataset assessment and
label acquisition.}

\maketitle

\clearpage

\section{Introduction}

Machine learning accelerates scientific discovery by enabling rapid property
prediction for candidate materials, molecules and reactions, reducing reliance
on costly experiments or first-principles calculations
\cite{morganOpportunitiesChallengesMachine2020,
butlerMachineLearningMolecular2018, ramprasadMachineLearningMaterials2017,
xiaoHighthroughputComputationalScreening2025}. These predictions inform
candidate selection in high-throughput screening and guide candidate generation
in emerging generative approaches \cite{xiaoAccurateScreeningFunctional2026,
takaharaAcceleratedInorganicMaterials2025,
hanInvDesFlowALActiveLearningbased2025, parkGuidingGenerativeModels2026}. As
these approaches explore new chemistries, their success depends on whether
models remain reliable when applied to chemistries not represented in their
training data. However, recent studies show that strong performance on conventional
random-split benchmarks can coexist with substantial degradation for unfamiliar
chemistries \cite{tenorioOutofdistributionMachineLearning2026,
liProbingOutofdistributionGeneralization2025,
omeeStructurebasedOutofdistributionOOD2024}. This challenge is especially
important for functional properties, for which costly measurements or
calculations often produce small datasets with uneven chemical coverage
\cite{xuSmallDataMachine2023}.

Existing studies address related aspects of chemical extrapolation from
different perspectives. Li \emph{et al.} showed that unfamiliar compounds in
poorly covered regions of materials-representation space generally exhibit
larger prediction errors, linking extrapolation difficulty to the coverage of
the labeled training data \cite{liProbingOutofdistributionGeneralization2025}.
Influence functions and data-valuation methods can estimate the contributions of
individual training samples, while uncertainty-based acquisition methods
identify additional samples for labeling \cite{pmlr-v70-koh17a,
pmlr-v97-ghorbani19c, liuUnderstandingImportanceIndividual2024,
smithANI1ccxANI1xData2020, liExploitingRedundancyLarge2023,
tanSinglemodelUncertaintyQuantification2023}. These approaches, however, depend
on target-property labels and typically on an already trained model. Chemical
similarity, by contrast, relates elements independently of the property being
predicted. Elemental properties and periodic-table coordinates are supplied as
input features that relate an element absent from labeled training to the
elements present in it \cite{mahdaviTrainingDataHow2025,
zhaoElementPriorsTarget2026}. Learned elemental embeddings are often interpreted
through their agreement with periodic trends
\cite{chenGraphNetworksUniversal2019,
onwuliElementSimilarityHighdimensional2023,
liPlatonicRepresentationFoundation2026}. It remains unclear which chemistries in
the training data inform predictions for an unfamiliar chemistry, and whether
chemical similarity or other relationships between elements can identify them.

\begin{figure}[h]
\centering
\includegraphics[width=0.98\textwidth]{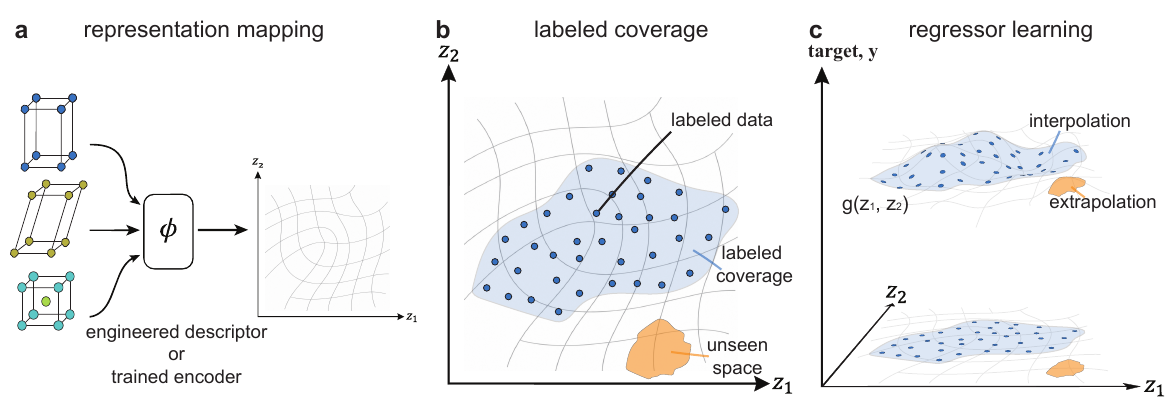}
\caption{\textbf{Representation mapping, labeled coverage and regressor learning.}
\textbf{a}, A representation mapping $\phi$, either an engineered descriptor or a
trained encoder, maps a material $x$ in materials space $\mathcal{X}$ to a point
$z=\phi(x)$ in representation space $\mathcal{Z}$, shown schematically with two
coordinates $z_1$ and $z_2$.
\textbf{b}, Labeled training data (points) cover part of $\mathcal{Z}$ (blue
region). Compounds from an unfamiliar chemistry may lie outside this labeled
coverage (orange region).
\textbf{c}, A downstream regressor $g$ learns from the labeled data, giving
predictions $\hat{y}=g(\phi(x))$. Predictions within the labeled coverage are
interpolations, whereas those outside it are extrapolations.}
\label{fig:representation}
\end{figure}

Training supervised property models involves three conceptually distinct
components: representation mapping, labeled coverage and regressor learning
(Fig.~\ref{fig:representation}) \cite{bengioRepresentationLearningReview2013}.
Because each labeled compound enters supervised learning as a pair of
representation and label, the representation mapping and labeled coverage are
often considered together. Conceptually, however, they play distinct roles. A
mapping $\phi$ determines how compounds are organized in representation space,
whereas the labeled dataset determines which regions in that space are paired
with target-property values and exposed to downstream learning. Because a
regressor $g$ learns representation--property relationships only from these
labeled pairs, its predictions for a target chemistry absent from the labeled
coverage rely entirely on pairs from other chemistries. Here we ask which
chemically defined subsets of the labeled training data expose
representation--property relationships most transferable to a given target
chemistry, and whether these subsets can be identified from the representation
mapping alone, without target-property labels or a trained regressor.

We define target chemistries using the \gls{loeo} protocol, in which compounds
containing a held-out element $X$ are excluded from training and serve as the
target set, while the remaining $X$-free compounds form the training pool
\cite{liProbingOutofdistributionGeneralization2025,
witmanMatFoldSystematicInsights2025}. We then divide the pool
into overlapping source subsets, each comprising the compounds that contain a
given element $Y$. Targets and sources are thus both indexed by elements, so
that each source subset represents a distinct chemistry and each target--source
pair corresponds to an element pair $(X,Y)$. Because both are defined by
composition alone, they are identical for every representation and model. We
compare two widely used classes of materials representations that differ in how
they are constructed. Engineered descriptors are built from predefined
elemental properties and operations, whereas pretrained crystal-graph encoders
learn representations through training. As representative examples, we use
Matminer descriptors and the encoder of the pretrained interatomic potential
eSEN-30M-OAM \cite{wardGeneralpurposeMachineLearning2016,
wardIncludingCrystalStructure2017,Ward2018Matminer,MatminerSoftware,
BarrosoLuque2024OMat24,fuLearningSmoothExpressive2025,OMat24_HuggingFace}. We
evaluate formation energy and bandgap, representing thermodynamic and electronic
properties, across JARVIS, MP and OQMD databases
\cite{jainCommentaryMaterialsProject2013a,
saalMaterialsDesignDiscovery2013a,kirklinOpenQuantumMaterials2015a,
choudharyJointAutomatedRepository2020,
hortonAcceleratedDatadrivenMaterials2025}.

We find that materials representations rank source subsets by the transferable
supervision they provide. Excluding top-ranked subsets at fixed training-set
size substantially increases extrapolation error, whereas excluding lower-ranked
subsets has little effect, and the rankings hold across properties, regressors
and datasets. The \gls{engineered} ranks source subsets largely by
elemental-property similarity, whereas the \gls{pretrained} ranks them by
directional alignment, which compares how compounds shift in representation
space as the fractions of two elements increase. This alignment identifies
useful subsets more effectively than elemental-property similarity or
periodic-group membership.

\section{Results}

\subsection{Element fraction tracks representation-space separation and extrapolation difficulty}
\label{sec:fraction}

\begin{figure}[h]
\centering
\includegraphics[width=0.98\textwidth]{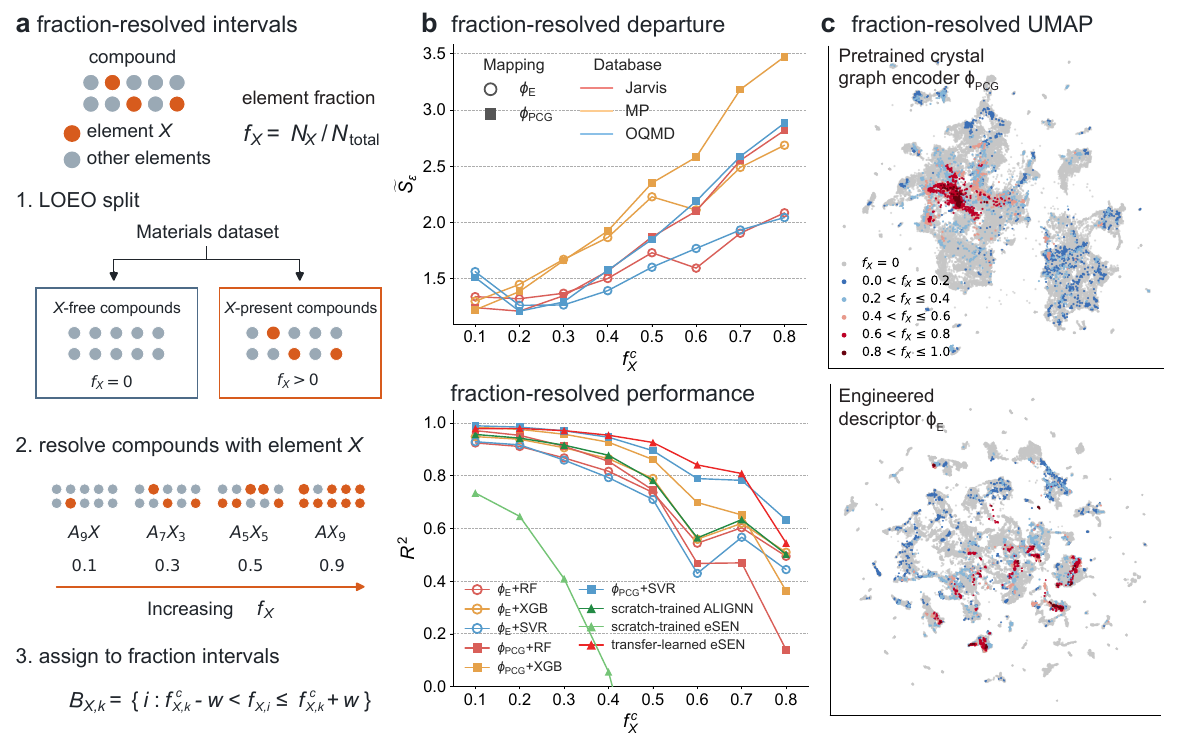}
\caption{\textbf{Element fraction resolves representation-space separation and
extrapolation difficulty.}
\textbf{a}, Fraction-resolved analysis. For each element $X$, compounds are
divided into $X$-free ($f_X=0$) and $X$-present ($f_X>0$) sets, and $X$-present
compounds are assigned to intervals $B_{X,k}$ of the element fraction
$f_X=N_X/N_{\mathrm{total}}$, with center $f_{X,k}^{\mathrm{c}}$ and half-width $w$.
\textbf{b}, Top, separation between the representation-space distribution of
each fraction interval and that of $X$-free compounds, quantified by the
normalized Sinkhorn divergence $\widetilde{S}_{\varepsilon}$ as a function of the
interval center \gls{efc}. The divergence is computed from the
representations without dimensionality reduction or property labels. Marker shapes distinguish
\glsentryshort{engineered} and \glsentryshort{pretrained}; colors distinguish
JARVIS, MP and OQMD. Bottom, formation-energy extrapolation performance on JARVIS
in a \glsentryshort{loeo} setting, in which models are trained on samples from
the $X$-free training pool and evaluated by $R^2$ on $X$-present target compounds
within each fraction interval. Fixed representations paired with RF, XGB and
SVR are compared with scratch-trained
ALIGNN, scratch-trained eSEN and transfer-learned eSEN. Intervals have $w=0.1$,
so adjacent intervals overlap. Values are means over eligible held-out elements
(64, 72 and 84 for JARVIS, MP and OQMD, respectively) and five
sampling repetitions.
\textbf{c}, Two-dimensional \glsentryshort{umap} projections of JARVIS compounds
in \glsentryshort{pretrained} (top) and \glsentryshort{engineered} (bottom).
Mg-free compounds ($f_{\mathrm{Mg}}=0$) are shown in gray, and Mg-present
compounds are colored by $f_{\mathrm{Mg}}$.}
\label{fig:fraction}
\end{figure}

Compounds containing a given chemical element $X$ ($=$~H, He, Li, \ldots)
can differ substantially in its atomic fraction, $f_X=N_X/N_{\mathrm{total}}$,
where $N_X$ is the number of $X$ atoms and $N_{\mathrm{total}}$ is the total
number of atoms in the compound. To examine how this variation relates to
representation-space separation and extrapolation performance, we grouped these
compounds into fraction intervals $B_{X,k}$ with centers $f_{X,k}^{\mathrm{c}}$
and half-width $w$ (Fig.~\ref{fig:fraction}a).

We first examined how the representation-space separation between each
fraction-interval distribution and the $f_X=0$ reference distribution varies
with the interval center \gls{efc} (Fig.~\ref{fig:fraction}b). We quantified
separation using \gls{nsinkhorn}, obtained by normalizing the Sinkhorn
divergence by a reference divergence within the $f_X=0$ distribution. For each
fraction interval, we report the mean \gls{nsinkhorn} across elements.
Separation generally increased with \gls{efc} in both \gls{engineered} and
\gls{pretrained}, indicating increasing departure from the $f_X=0$ distribution.

We then evaluated formation-energy extrapolation toward different fraction
intervals (Fig.~\ref{fig:fraction}b). Each eligible element was used as a
held-out element $X$ in turn. Models were trained on samples from the $X$-free
training pool and evaluated on $X$-present target compounds within each
interval. Besides pairing the
\gls{engineered} and \gls{pretrained} representations with \gls{rf}, \gls{xgb}
and \gls{svr}, we also evaluated scratch-trained ALIGNN, scratch-trained eSEN
and transfer-learned eSEN on the JARVIS formation-energy dataset
\cite{choudharyAtomisticLineGraph2021, xiaoAccurateScreeningFunctional2026,
ALIGNN, MLIP_FTL}. For each fraction interval, we report the mean $R^2$ across
held-out elements. Most models approached the interpolation performance at low
\gls{efc} \cite{liProbingOutofdistributionGeneralization2025}. As \gls{efc}
increased, performance generally declined and differences among models became
more pronounced, revealing different extrapolation behaviors. This
trend was observed across JARVIS, MP and OQMD (Fig.~\ref{fig:fraction}b and
Supplementary Figs.~1 and 2). Extrapolation difficulty therefore varies among
the target compounds of a held-out element, with higher $f_X$ generally posing a
greater challenge.

Resolving compounds by $f_X$ also offers a compositional view of representation
overlap. Li \emph{et al.} found that regions with little overlap between
$X$-present and $X$-free compounds were concentrated in \gls{umap} projections
of learned encoders but dispersed in those of engineered descriptors
\cite{liProbingOutofdistributionGeneralization2025}. Using Mg-present compounds
as an example (Fig.~\ref{fig:fraction}c), low-$f_{\mathrm{Mg}}$ compounds were
interspersed with Mg-free compounds in both projections. With increasing
$f_{\mathrm{Mg}}$, Mg-present compounds followed a concentrated progression in
\gls{pretrained} but extended along multiple dispersed branches in
\gls{engineered}, mirroring the reported distribution of low-overlap regions.
Together with the rise in separation and extrapolation difficulty with $f_X$
(Fig.~\ref{fig:fraction}b), this suggests that low-overlap regions consist
largely of compounds with high $f_X$. The element fraction thus provides a
representation-independent coordinate for interpreting overlap between
$X$-present and $X$-free compounds in the representation space.

\subsection{Directional alignment identifies useful source subsets in \gls{pretrained}}
\label{sec:pretrain}

In the original representation space, compositional trajectories connecting the
successive fraction intervals were approximately linear in \gls{pretrained}
(Supplementary Figs.~3 and 4), consistent with the concentrated progression of
Mg-present compounds in the \gls{umap} projection (Fig.~\ref{fig:fraction}c).
This linearity allows the change in representation with increasing $f_X$ to be
summarized by a direction vector for each element. We quantify alignment between
the directions of different elements and show that it identifies source subsets
that provide transferable supervision.

For each element $X$, we used five fraction intervals with centers
$f^{\mathrm c}_{X,k}\in\{0.1,0.3,0.5,0.7,0.9\}$ and half-width
$w=0.1$. We fitted a line to their centroids in the original
representation space to obtain a compositional direction,
\begin{equation}
\boldsymbol{\mu}_{X,k}
=
\mathbf{a}_X
+
f^{\mathrm c}_{X,k}\mathbf{v}_X
+
\boldsymbol{\epsilon}_{X,k},
\label{eq:trajectory_vector}
\end{equation}
where $\boldsymbol{\mu}_{X,k}$ and $f^{\mathrm c}_{X,k}$ are the representation centroid and
the center of interval $k$; $\mathbf{v}_X$, $\mathbf{a}_X$ and
$\boldsymbol{\epsilon}_{X,k}$ are the fitted direction vector, intercept and
residual, respectively. Directional alignment between elements $X$ and $Y$ is
defined as the cosine similarity of their fitted direction vectors,
\begin{equation}
A_{XY}
=
\frac{
\mathbf{v}_X^{\mathrm T}\mathbf{v}_Y
}{
\lVert\mathbf{v}_X\rVert_2
\lVert\mathbf{v}_Y\rVert_2
}.
\label{eq:directional_alignment}
\end{equation}
Values approaching one indicate similarly directed changes in representation
space with increasing $X$ or $Y$ content. We denote the alignment computed in
\gls{pretrained} by $A_{XY}^{\mathrm{PCG}}$, which can be interpreted as a
geometric similarity between elements in the learned compound representation.

Computing $A_{XY}$ requires no target-property labels, but it is influenced by
the collection of compounds used as probes in representation space. To
assess this dependence, we ranked, for each element $X$, the other elements by
$A_{XY}^{\mathrm{PCG}}$ computed separately from JARVIS, MP and OQMD, and
compared the rankings between datasets using the Spearman rank correlation
coefficient $\rho$. The rankings were highly consistent, with element-averaged
$\rho$ values of 0.92 for JARVIS--MP, 0.83 for JARVIS--OQMD and 0.82 for
MP--OQMD.

To test whether $A_{XY}^{\mathrm{PCG}}$ identifies source subsets that provide
transferable supervision, we performed a source-exclusion analysis, removing
selected source subsets from the training pool and measuring the resulting
change in extrapolation \gls{mae}. For each held-out element $X$, source subsets
were ranked by $A_{XY}^{\mathrm{PCG}}$ computed from OQMD, the largest of the
three datasets, and this single ranking was used for both properties and for
every regressor and dataset. Starting from a baseline pool of all $X$-free
compounds, we removed the top two, four or six source subsets or, as a control,
six lower-ranked subsets (ranks 11--16). Models were trained on 10,000 compounds
sampled from each pool, so that conditions differed in chemical coverage but not
in training-set size. Test sets consisted of target compounds with $f_X>0.4$,
for which extrapolation is more challenging.

\begin{figure}[h!]
\centering
\includegraphics[width=0.99\textwidth]{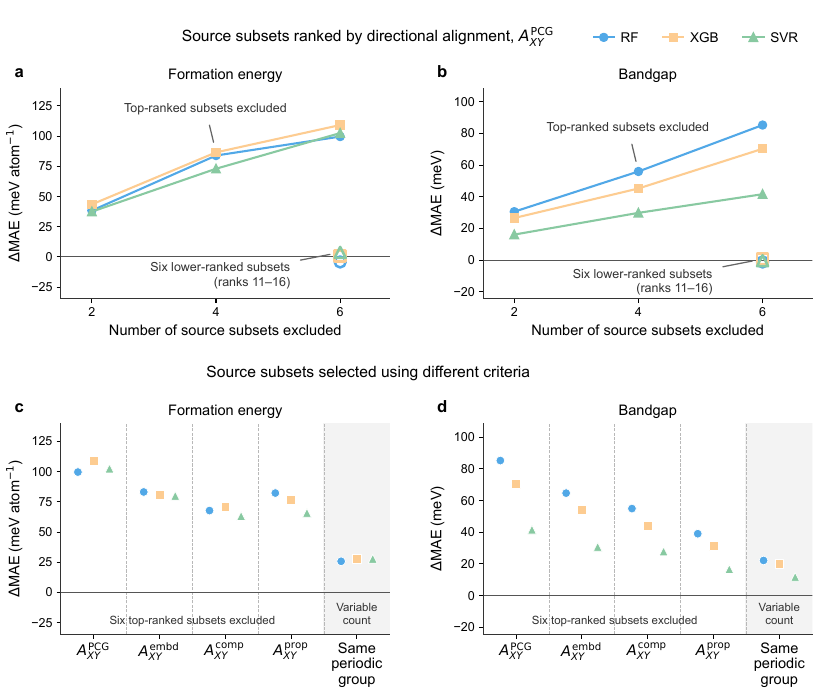}
\caption{\textbf{Directional alignment in the pretrained representation identifies
useful source subsets.}
Changes in \gls{mae} ($\Delta$MAE) after source exclusion for formation-energy
(\textbf{a},\textbf{c}) and bandgap (\textbf{b},\textbf{d}) prediction on OQMD
using the \glsentryshort{pretrained} representation, relative to a baseline
trained on samples from the full $X$-free training pool. A source subset
comprises compounds containing a source element $Y$ but not the held-out element
$X$.
\textbf{a,b}, Removal of the top two, four or six source subsets from the
training pool, ranked by $A_{XY}^{\mathrm{PCG}}$ (filled markers and lines),
and of six lower-ranked subsets (ranks 11--16; hollow markers).
\textbf{c,d}, Top-six exclusion with source subsets ranked by alignment in the
final compound representation ($A_{XY}^{\mathrm{PCG}}$), elemental-embedding
similarity ($A_{XY}^{\mathrm{embd}}$), alignment before message passing
($A_{XY}^{\mathrm{comp}}$) and elemental-property similarity
($A_{XY}^{\mathrm{prop}}$). The shaded column shows exclusion of source subsets
whose source elements share the periodic group of $X$; their number varies with
$X$. All conditions use 10,000 training compounds. Test sets contain target
compounds with $f_X>0.4$, filtering out compounds that contain any of the six
highest-ranked source elements of the ranking measure used; each condition and
its baseline share identical test compounds. Each marker is the mean of
$\Delta$MAE pooled over 61 held-out elements and five sampling repetitions after
discarding the lowest and highest 10\% of values. Colors and marker shapes
distinguish RF, XGB and SVR. Results on JARVIS and MP are shown in Supplementary
Figs.~5 and 6, all values in Supplementary Figs.~7--9, and results with
unfiltered test sets in Supplementary Figs.~10--12.}
\label{fig:directional}
\end{figure}

\gls{mae} increased progressively as the number of excluded top-ranked source
subsets rose from two to four to six, whereas ranks 11--16 exclusion produced
little change (Fig.~\ref{fig:directional}a,b). On OQMD, top-six exclusion
increased MAE by approximately $100$--$110$~meV atom$^{-1}$ for formation energy
and $42$--$85$~meV for bandgap across the three regressors, compared with
changes of less than 5~meV atom$^{-1}$ and 2~meV for ranks 11--16 exclusion. The
same pattern held on JARVIS and MP for every regressor and property
(Supplementary Figs.~5--12). Ranking by $A_{XY}^{\mathrm{PCG}}$ therefore identifies source
subsets whose labeled examples provide transferable supervision for the target
compounds. This single ranking, computed from OQMD without target-property
labels, remained informative across properties, regressors and datasets,
suggesting that the representation mapping largely determines which source
subsets are useful.

We next asked whether $A_{XY}^{\mathrm{PCG}}$ simply reflects conventional
chemical similarity between elements. We compared it with two such measures, one
based on membership in the same periodic group as $X$ and the other on
$A_{XY}^{\mathrm{prop}}$, the cosine similarity between 22-dimensional vectors
of elemental properties such as electronegativity and covalent radius. Rankings
by $A_{XY}^{\mathrm{PCG}}$ and $A_{XY}^{\mathrm{prop}}$ agreed only moderately
($\rho=0.46$). Excluding the top six source subsets ranked by
$A_{XY}^{\mathrm{prop}}$ increased MAE, but less than excluding the top six
ranked by $A_{XY}^{\mathrm{PCG}}$, and the same-group control, whose exclusion
count varies with $X$, also produced smaller increases
(Fig.~\ref{fig:directional}c,d). Alignment in the \gls{pretrained}
representation therefore departs from elemental-property similarity and
identifies useful source subsets more effectively.

\subsection{Message passing refines relationships captured by $A_{XY}$}
\label{sec:message}

In \gls{pretrained}, each atom is first represented by a learned embedding of
its element; these atomic features are then updated through message-passing
layers and averaged over all atoms to obtain the crystal representation
\cite{fuLearningSmoothExpressive2025,
liProbingOutofdistributionGeneralization2025, kimLeveragingNeuralNetwork2026,
SugiuraAchievingRobust2026}. To determine whether the ranking captured by
$A_{XY}^{\mathrm{PCG}}$ is already present before message passing, we examined
two measures derived from the elemental embeddings of \gls{pretrained}.
$A_{XY}^{\mathrm{embd}}$ is the cosine similarity between the embedding vectors
of elements $X$ and $Y$. $A_{XY}^{\mathrm{comp}}$ is the directional alignment
of compositional trajectories fitted to compound representations formed by
averaging the element embeddings according to atomic fractions, that is, the
representations before message passing.

Rankings by $A_{XY}^{\mathrm{embd}}$ were moderately correlated with those by
$A_{XY}^{\mathrm{prop}}$ ($\rho=0.42$), suggesting that \gls{pretrained} already
departs from conventional chemical similarity at the embedding level. Message
passing then substantially changed relationships in embeddings, as rankings by
$A_{XY}^{\mathrm{PCG}}$ were only moderately correlated with those by
$A_{XY}^{\mathrm{comp}}$ ($\rho=0.55$) and $A_{XY}^{\mathrm{embd}}$
($\rho=0.48$). For example, the six source elements ranked highest for Cu by
$A^{\mathrm{embd}}_{XY}$ (Ni, Au, Pt, Fe, V and Mn) include Fe, V and Mn from
groups 5--8, whereas those ranked highest by $A^{\mathrm{PCG}}_{XY}$ (Ag, Au,
Ni, Hg, Cd and Pd) all belong to groups 10--12 (Supplementary Table~1). Excluding
the six source subsets ranked highest by $A_{XY}^{\mathrm{embd}}$ or
$A_{XY}^{\mathrm{comp}}$ increased MAE, but less than excluding those ranked
highest by $A_{XY}^{\mathrm{PCG}}$, for every combination of property, regressor
and dataset (Fig.~\ref{fig:directional}c,d and Supplementary Figs.~5 and 6). The
learned elemental embeddings thus provide a useful starting point, but the most
informative ranking emerges once message passing places each element in its
chemical environment.

Besides changing rankings, message passing also affects the linear geometry of
the trajectories on which $A_{XY}$ relies. Because message passing makes each
atomic feature depend on its local chemical environment, atoms of the same
element can have different features, and linear compositional trajectories are
not guaranteed by mean pooling alone. In \gls{pretrained}, trajectories became
progressively less linear over the first message-passing layers but remained
approximately linear at the fourth layer used here (Supplementary Fig.~4). Li
and Walsh found that atomic representations after message passing are still
organized by element across foundation interatomic potentials with different
architectures and training data~\cite{liPlatonicRepresentationFoundation2026}.
Under mean pooling, such element-dependent organization suggests that the
observed linearity is common to encoders extracted from pretrained interatomic
potentials.

\subsection{Elemental-property similarity identifies useful source
subsets in \gls{engineered}}
\label{sec:engineer}

Like \gls{pretrained}, \gls{engineered} can be viewed as a set of operations
applied to elemental embeddings, but in \gls{engineered} both are predefined.
Most of its descriptors are statistics of 22 elemental properties, such as
electronegativity and covalent radius, pooled over the constituent elements or
over property differences between Voronoi neighbors
\cite{wardGeneralpurposeMachineLearning2016, wardIncludingCrystalStructure2017,
Ward2018Matminer}. The 22-dimensional property vector of each element therefore
acts as a fixed elemental embedding, and neighbor differences take the place of
message passing. The similarity between elemental-property vectors,
$A_{XY}^{\mathrm{prop}}$, which served as a control for \gls{pretrained}
(Fig.~\ref{fig:directional}), is thus the counterpart of the embedding
similarity $A_{XY}^{\mathrm{embd}}$ in \gls{engineered}. Of the pooling
statistics, only the fraction-weighted means vary linearly with atomic fraction,
so trajectories in \gls{engineered} were much less linear than in
\gls{pretrained} (Supplementary Fig.~3).

\begin{figure}[h]
\centering
\includegraphics[width=0.95\textwidth]{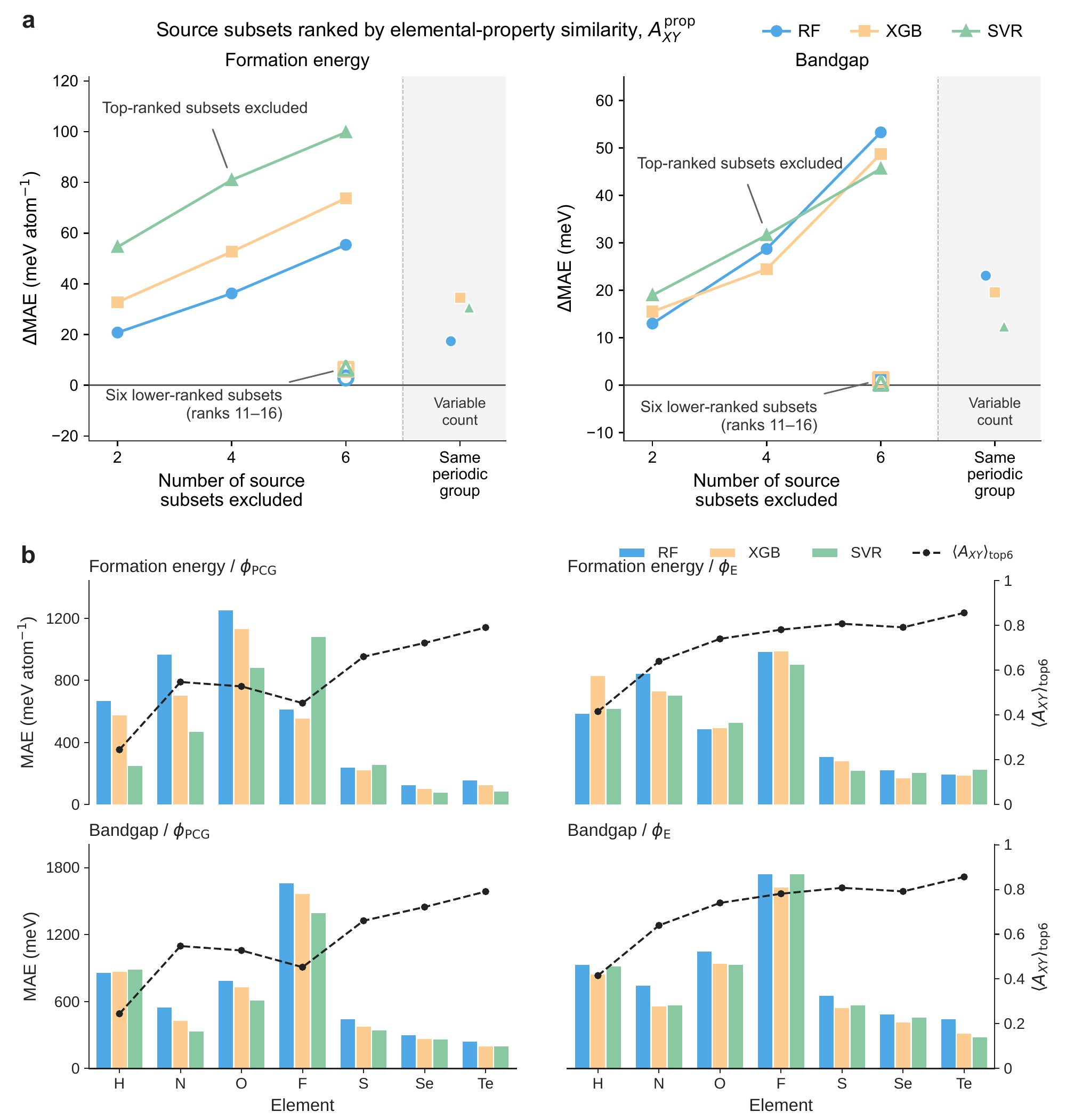}
\caption{\textbf{Source exclusion with engineered descriptors and
element-specific extrapolation error.} \textbf{a}, $\Delta$MAE after source
exclusion for formation-energy (left) and bandgap (right) prediction on OQMD
using the fixed \glsentryshort{engineered} representation. Filled markers and
lines show removal of the top two, four or six source subsets from the training
pool, ranked by elemental-property similarity, $A_{XY}^{\mathrm{prop}}$; hollow
markers show exclusion of six lower-ranked subsets (ranks 11--16). The shaded
column shows the same-periodic-group control, whose exclusion count varies with
$X$. Results on JARVIS and MP are shown in Supplementary Figs.~13 and 14, all
values in Supplementary Figs.~15--17, and results with unfiltered test sets in
Supplementary Figs.~18--20.
\textbf{b}, Extrapolation MAE on target compounds for held-out H, N, O, F, S, Se
and Te on OQMD, for formation energy (top) and bandgap (bottom) with
\glsentryshort{pretrained} (left) and \glsentryshort{engineered} (right). Models
were trained on 10,000 compounds sampled from each full training pool without
source exclusion. Colors distinguish RF, XGB and SVR. The dashed line (right
axis) shows the top-six score $\langle A_{XY}\rangle_{\mathrm{top6}}$, the mean
of the six largest values of $A_{XY}^{\mathrm{PCG}}$ (left) or
$A_{XY}^{\mathrm{prop}}$ (right). Results on JARVIS and MP are shown in
Supplementary Figs.~21 and 22.}
\label{fig:Engineer}
\end{figure}

We tested whether $A_{XY}^{\mathrm{prop}}$ identifies useful source subsets by
repeating the exclusion study with \gls{engineered}. MAE increased as the number
of excluded top-ranked source subsets rose, whereas excluding six lower-ranked
subsets (ranks 11--16) produced small changes (Fig.~\ref{fig:Engineer}a). On
OQMD, top-six exclusion increased MAE by approximately $55$--$100$~meV
atom$^{-1}$ for formation energy and $46$--$53$~meV for bandgap across the three
regressors, whereas exclusion of ranks 11--16 changed MAE by less than 7~meV
atom$^{-1}$ and 2~meV, respectively. The same graded pattern, with little change
for ranks 11--16, was observed across all evaluated regressors, target
properties and datasets (Supplementary Figs.~13--20).

In \gls{engineered}, periodic-group membership was nearly as informative as
elemental-property similarity.
The same-group control removed two to five source subsets. With RF and XGB, its
MAE increases were comparable to those of top-two to top-four exclusion by
$A_{XY}^{\mathrm{prop}}$, whereas with SVR they were smaller than those of
top-two exclusion (Fig.~\ref{fig:Engineer}a).
This agreement is expected, because the periodic
table reflects trends in elemental properties, making group membership a coarse
proxy for the elemental-property similarity on which \gls{engineered} is built.
In \gls{pretrained}, by contrast, same-group exclusion produced smaller
increases than even top-two exclusion by $A_{XY}^{\mathrm{PCG}}$ for every
combination of property, regressor and dataset (Fig.~\ref{fig:directional} and
Supplementary Figs.~5--9).

\subsection{Top-ranked $A_{XY}$ values indicate element-specific extrapolation
difficulty}

Beyond ranking source subsets for a given held-out element, we asked whether
$A_{XY}$ values also indicate which held-out elements are difficult to
extrapolate to. For each held-out element $X$, we defined the top-six score
$\langle A_{XY}\rangle_{\mathrm{top6}}$ as the mean of the six largest $A_{XY}$
values, using $A_{XY}^{\mathrm{PCG}}$ for \gls{pretrained} and
$A_{XY}^{\mathrm{prop}}$ for \gls{engineered}. A low score indicates that even
the top-ranked source elements relate only weakly to $X$. We compared this score
with the extrapolation MAE obtained without source exclusion. Across held-out
elements, higher scores were associated with lower MAE for every combination of
representation, regressor, property and dataset, with Spearman $\rho$ ranging
from $-0.74$ to $-0.29$ (Table~\ref{tab:support_error_correlations}). The two
representations gave similar correlations for formation energy ($-0.52$ to
$-0.29$), whereas for bandgap, correlations were stronger with \gls{pretrained}
($-0.74$ to $-0.65$) than with \gls{engineered} ($-0.58$ to $-0.42$) in every
combination of dataset and regressor. The moderate strength of these
correlations is expected, because extrapolation error also depends on factors
that the score does not capture, such as the labeled coverage within useful
source subsets and how effectively the regressor learns from them.

\begin{table}[t]
\caption{Spearman rank correlations between top-six scores and
chemical extrapolation error.}
\label{tab:support_error_correlations}
\centering
\begin{tabular*}{\linewidth}{@{\extracolsep{\fill}}llrrrr@{}}
\toprule
 & & \multicolumn{2}{c}{Formation energy} & \multicolumn{2}{c}{Bandgap} \\
\cmidrule(lr){3-4}\cmidrule(lr){5-6}
Dataset & Regressor & $\phi_{\mathrm{PCG}}$ & $\phi_{\mathrm E}$
 & $\phi_{\mathrm{PCG}}$ & $\phi_{\mathrm E}$ \\
\midrule
JARVIS & RF & $-0.46$ & $-0.44$ & $-0.65$ & $-0.44$ \\
 & XGB & $-0.52$ & $-0.49$ & $-0.70$ & $-0.46$ \\
 & SVR & $-0.42$ & $-0.48$ & $-0.74$ & $-0.58$ \\
\addlinespace
MP & RF & $-0.40$ & $-0.30$ & $-0.67$ & $-0.42$ \\
 & XGB & $-0.44$ & $-0.39$ & $-0.71$ & $-0.47$ \\
 & SVR & $-0.33$ & $-0.29$ & $-0.73$ & $-0.56$ \\
\addlinespace
OQMD & RF & $-0.39$ & $-0.41$ & $-0.70$ & $-0.45$ \\
 & XGB & $-0.45$ & $-0.41$ & $-0.70$ & $-0.51$ \\
 & SVR & $-0.38$ & $-0.44$ & $-0.70$ & $-0.56$ \\
\bottomrule
\end{tabular*}
\par\smallskip
\begin{minipage}{\linewidth}
\footnotesize
Coefficients are Spearman $\rho$ calculated across held-out elements between
the top-six score, $\langle A_{XY}\rangle_{\mathrm{top6}}$, and extrapolation
MAE. The top-six score is the mean of the six largest values of
$A_{XY}$ over source elements $Y$ indexing the source subsets, using
$A_{XY}^{\mathrm{PCG}}$ for the pretrained crystal-graph encoder
($\phi_{\mathrm{PCG}}$) and $A_{XY}^{\mathrm{prop}}$ for the engineered
descriptors ($\phi_{\mathrm E}$). $A_{XY}^{\mathrm{PCG}}$ was computed from
OQMD for all three datasets. MAE was evaluated on filtered test sets of
target compounds with $f_X>0.4$. Negative coefficients indicate
that higher top-six scores are associated with lower error. RF, random forest;
XGB, extreme gradient boosting; SVR, support-vector regression.
\end{minipage}
\end{table}

To illustrate this relationship for individual elements, we examined H, N, O and
F, for which previous studies have reported particularly poor extrapolation
\cite{witmanMatFoldSystematicInsights2025,
liProbingOutofdistributionGeneralization2025,
vasconcelosInterpretableFormationenergyStructure2026}. In \gls{pretrained},
these elements had lower $\langle A_{XY}\rangle_{\mathrm{top6}}$ and generally
higher extrapolation MAEs than the heavier chalcogens S, Se and Te
(Fig.~\ref{fig:Engineer}b and Supplementary Figs.~21 and 22). In
\gls{engineered}, the correspondence was weaker, as elemental-property
similarity gave O and F relatively high scores despite their poor extrapolation.
For these elements, the element similarity learned by \gls{pretrained} therefore
tracks extrapolation difficulty more closely than elemental-property similarity
does. Within H, N, O and F, however, the scores did not follow the order of the
errors in either representation, indicating that representation-level measures
alone cannot fully account for extrapolation difficulty.

\section{Discussion}

Our results show that materials representations can identify which source
subsets of a training pool supply transferable supervision for target compounds,
without requiring target-property labels. A single ranking for each held-out element remained
informative across properties, regressors and datasets, indicating that which
subsets are useful is set largely by the representation. These findings clarify
the distinct roles that the three components in Fig.~\ref{fig:representation}
play in chemical extrapolation. The representation determines which compounds
expose representation--property relationships that carry over to the
extrapolation, labeled coverage determines whether labels of such compounds are
available, and the regressor determines how effectively these relationships are
learned. In this sense, a representation organizes supervision.

The two representations organized this supervision differently. In
\gls{engineered}, useful source subsets largely followed elemental-property
similarity and periodic-group membership, as expected for descriptors built from
predefined elemental properties. In \gls{pretrained}, the learned element
similarity differed substantially from both criteria and identified useful
source subsets more effectively than either. Unlike \gls{engineered},
\gls{pretrained} receives no elemental properties as input. Its elemental
embeddings are learned entirely during pretraining on compounds, and message
passing further makes atomic features depend on local chemical environments. The
difference may therefore arise because elemental properties characterize
elements in isolation, whereas \gls{pretrained} relates elements by how they
behave within compounds.

Because the rankings require no target-property labels, they separate which
source subsets can supply supervision from whether labeled data cover those
subsets. This separation matters most when labels are scarce, as for many
functional materials whose property datasets are small and chemically uneven
\cite{xuSmallDataMachine2023}. Rankings computed from OQMD remained informative
when applied to JARVIS and MP, so for a fixed representation they can be
estimated once from a large collection of probe compounds and reused.
A small labeled dataset could then be assessed by its coverage of the source
subsets ranked highly for a target chemistry.
Because probe compounds serve only to sample representation space, rankings
computed from first-principles databases could also be applied to datasets with
experimental labels.
Poor extrapolation could likewise be traced to one of two causes. If even the
top-ranked source subsets relate only weakly to the target chemistry, which is
associated with higher extrapolation error
(Table~\ref{tab:support_error_correlations}), a different representation may be
needed. If highly ranked subsets instead lack labeled coverage, labels could be
acquired within them, complementing acquisition criteria based on model
uncertainty or data diversity \cite{tanSinglemodelUncertaintyQuantification2023,
smithANI1ccxANI1xData2020, liExploitingRedundancyLarge2023}.

More broadly, our results suggest that the value of labeled data for chemical
extrapolation is determined not by the data alone but also by the representation
through which they are learned. Which labeled compounds are most informative for
a new chemistry can differ between representations, and this can be assessed
before any property model is trained. Evaluating materials representations by
how they organize supervision, alongside their predictive accuracy, could
therefore guide both the choice of representation and the design of datasets for
exploring new chemistries.

\section{Methods}

\subsection{Datasets, representations and regression models}

We used the JARVIS22, MP21 and OQMD21 datasets (hereafter JARVIS, MP and OQMD)
from the Zenodo archive accompanying the study by Li \emph{et
al.}~\cite{liExploitingRedundancyLarge2023}. The engineered representation
\gls{engineered} consisted of the 273 Matminer descriptors precomputed for each
compound and distributed with the same archive~\cite{Ward2018Matminer,
liExploitingRedundancyLarge2023}. Of these, 242 are statistics of 22 elemental
properties, such as electronegativity, covalent radius and valence-electron
counts, with 132 computed over the constituent elements of a compound
(fraction-weighted mean, mean absolute deviation, minimum, maximum, range and
mode), and 110 from property differences between neighboring atoms identified by
Voronoi tessellation~\cite{wardGeneralpurposeMachineLearning2016,
wardIncludingCrystalStructure2017}. A further six are derived from subsets of
these properties, and the remaining 25 describe stoichiometry, charge balance,
coordination, bond-length variation, packing efficiency and chemical ordering.
The pretrained representation \gls{pretrained}
was obtained from the embedding layer and the first four message-passing layers
of the pretrained eSEN-30M-OAM model~\cite{fuLearningSmoothExpressive2025,
BarrosoLuque2024OMat24, OMat24_HuggingFace}. For each crystal, we extracted the
rotationally invariant scalar ($\ell=0$) components of the node features after
the fourth message-passing layer and averaged them over all atoms, giving a
128-dimensional representation. The encoder weights were kept fixed throughout
this study. Both representations were standardized feature-wise to zero mean and
unit variance. For the geometric analyses, which use no target-property labels,
the means and standard deviations were computed from all compounds in each
dataset, which served as probes of representation space. For property
prediction, they were computed from the sampled
training set only.
Each representation was paired with three regressors, \gls{rf}, \gls{xgb} and
\gls{svr}. RF and XGB used the hyperparameters reported by Li \emph{et
al.}~\cite{liProbingOutofdistributionGeneralization2025}. SVR was implemented in cuML
with a radial basis function kernel, a regularization parameter $C=1.0$, an
$\epsilon$-insensitive loss with $\epsilon=0.1$ and the kernel coefficient set
by the `scale' heuristic \cite{raschkaMachineLearningPython2020}.
Hyperparameters were not tuned for individual representations, properties or
held-out elements.

\subsection{Fraction-resolved separation and performance}
\label{sec:method_fraction_resolved}

For each element $X$, $X$-present compounds ($f_X>0$) were grouped into
fraction intervals
\begin{equation}
B_{X,k}=\left\{
i:
f_{X,k}^{\mathrm{c}}-w
<
f_{X,i}
\leq
f_{X,k}^{\mathrm{c}}+w
\right\},
\end{equation}
where $f_{X,k}^{\mathrm{c}}$ is the center of interval $B_{X,k}$, $w$ is its
half-width, and $i$ indexes compounds. For the analyses in
Fig.~\ref{fig:fraction}b, we used eight overlapping intervals with centers
$f_X^{\mathrm{c}}=0.1,0.2,\ldots,0.8$ and half-width $w=0.1$. Elements were
analyzed only if every interval contained at least 50 compounds, giving 64, 72
and 84 elements for JARVIS, MP and OQMD, respectively.

Representation-space separation was quantified with the debiased Sinkhorn
divergence ($S_{\varepsilon}$) implemented in GeomLoss, using a quadratic cost ($p=2$) and a blur of
0.05, which corresponds to $\varepsilon=0.05^{2}$~\cite{pmlr-v89-feydy19a}. Let
$Z_{X,k}=\{\mathbf{z}_i: i\in B_{X,k}\}$ denote the representations of the
compounds in interval $B_{X,k}$, and $Z_{X,0}$ those of the $X$-free compounds.
The divergence was normalized by a reference divergence between two disjoint
subsamples of the $X$-free compounds, $Z_{X,0}^{(1)}$ and $Z_{X,0}^{(2)}$:
\begin{equation}
\widetilde{S}_{\varepsilon}(B_{X,k})
=
\frac{S_{\varepsilon}\bigl(Z_{X,k},\,Z_{X,0}\bigr)}
{S_{\varepsilon}\bigl(Z_{X,0}^{(1)},\,Z_{X,0}^{(2)}\bigr)}.
\end{equation}
A value of one therefore corresponds to the separation between two samples of
$X$-free compounds.

For predictive evaluation, training sets of 50,000, 70,000 and 100,000
compounds were sampled from the $X$-free training pools of JARVIS, MP and OQMD,
respectively, and models were evaluated on the $X$-present target compounds in
each interval. For each held-out element, the training set was sampled five times
with the test set fixed. $R^2$ was computed within each interval for each
repetition and then averaged over held-out elements and repetitions.

For comparison, three end-to-end models were trained on JARVIS formation
energies. ALIGNN used the configuration reported by Li \emph{et
al.}~\cite{liProbingOutofdistributionGeneralization2025}. The scratch-trained eSEN
model had four message-passing layers. The
transfer-learned eSEN model had six message-passing layers, of which the first
four were transferred from eSEN-30M-OAM and kept frozen, following the
frozen-transfer-learning approach implemented in
MLIP-FTL~\cite{MLIP_FTL}.

\subsection{Source-exclusion analysis}

In the source-exclusion analysis (Figs.~\ref{fig:directional}
and~\ref{fig:Engineer}a), the baseline training pool for a held-out element $X$
consisted of all $X$-free compounds. The source subset for element $Y$ comprised
the compounds in this pool containing $Y$, so its membership depended on both
$X$ and $Y$. Source subsets can overlap because a compound may contain several
source elements. Exclusion pools were formed by removing the union of the top
two, four or six source subsets ranked for $X$ by a given measure, or the union
of the six subsets ranked 11--16.
Rankings by $A_{XY}^{\mathrm{PCG}}$ and $A_{XY}^{\mathrm{comp}}$ were computed
from OQMD probe compounds and applied to source subsets in all three datasets;
rankings by
$A_{XY}^{\mathrm{embd}}$ and $A_{XY}^{\mathrm{prop}}$ do not depend on a
compound collection.
In the same-group control, source subsets whose source elements share the
periodic group of $X$ were removed. Five training sets of 10,000 compounds were
sampled from each pool, so that conditions differed in chemical coverage but not
in training-set size. Held-out elements were required to have at least 100 test
compounds. The 32 elements assigned to group 3 (Sc, Y, the lanthanides and the
actinides) were not used as held-out elements, because the same-group control
would then remove an unusually large number of source subsets. These criteria
yielded 60, 61 and 61 held-out elements for JARVIS, MP and OQMD, respectively.

Extrapolation was evaluated on target compounds with $f_X>0.4$ using two test
sets. The unfiltered set contained all such compounds, and the filtered set
excluded compounds containing any of the six source elements ranked highest by
the measure under study. Filtering removes a confound introduced by source
exclusion. When the source subset for $Y$ is removed from the training pool,
target compounds containing both $X$ and $Y$ require extrapolation to two
elements absent from the labeled training data. Changes in their errors cannot
then be attributed solely to the loss of transferable supervision from that
source subset. For each measure, the same filtered set was used for the top-two,
top-four, top-six and ranks 11--16 conditions. The same-group control used the
$A_{XY}^{\mathrm{PCG}}$ filter with \gls{pretrained} and the
$A_{XY}^{\mathrm{prop}}$ filter with \gls{engineered}. Because the filter
removes only compounds containing the six highest-ranked source elements, the
test sets for the ranks 11--16 and same-group controls can retain target
compounds that contain an element removed under these controls. Any resulting
double extrapolation would, if anything, increase the $\Delta$MAE of these
controls, making their comparison with top-ranked exclusion conservative. Each
exclusion condition was evaluated on the same test compounds as its baseline.
For each held-out element and repetition, we computed the change in \gls{mae},
$\Delta$MAE, relative to the baseline. $\Delta$MAE values were pooled across
held-out elements and repetitions and summarized by a trimmed mean, which
discards the lowest and highest values to limit the influence of extreme values.
The conclusions were insensitive to the trimming fraction, and we show results
obtained by discarding 10\% of values at each end. Results for the filtered test
sets are shown in the main text; those for the unfiltered test sets are given in
Supplementary Figs.~10--12 and 18--20 and lead to the same conclusions.

\backmatter

\bmhead{Data availability}

Data in this study are available at
\url{https://huggingface.co/datasets/nims-spin-theory/Axy_extrapolation}.

\bmhead{Code availability}

Scripts in this study are available at
\url{https://github.com/nims-spin-theory/Axy_extrapolation}.

\bmhead{Acknowledgements}
This research was conducted using the Supermicro ARS-111GL-DNHR-LCC and FUJITSU Server
PRIMERGY CX2550 M7 (Miyabi) at Joint Center for Advanced High Performance Computing
(JCAHPC) and the NIMS Numerical Materials Simulator.

\bmhead{Funding}
This study was partially supported by the MEXT Program: Data Creation and Utilization-Type
Material Research and Development Project (Digital Transformation Initiative
Center for Magnetic Materials), Grant Number JPMXP1122715503, and by JSPS KAKENHI Grant Numbers 25H01252 and 26K24734.

\bmhead{Author contributions}
T.T. conceptualized and designed the project. E.X. performed the calculations
and analyzed the data. Both authors contributed to the discussion and drafted
the manuscript.

\bmhead{Competing interests}
The authors declare no competing interests.

\bmhead{Supplementary information}
Supplementary Figures 1--22 and Supplementary Table 1 are provided in a separate
Supplementary Information file.

\bibliography{ref}

\end{document}